\documentclass[traditabstract]{aa}

\usepackage{graphicx}
\usepackage{txfonts}
\newcommand\kms{{\rm\,km\,s^{-1}}}
\newcommand\msun{\rm\,M_\odot}

\begin{document}

\title{Massive runaway stars in the Large Magellanic Cloud}

\author{V.V.~Gvaramadze\inst{1,2,3}
\and P.~Kroupa\inst{1} \and J.~Pflamm-Altenburg\inst{1}}

\institute{Argelander-Institut f\"{u}r Astronomie, Universit\"{a}t
Bonn, Auf dem H\"{u}gel 73, 53121 Bonn,
Germany\\\email{pavel@astro.uni-bonn.de (PK);
jpflamm@astro.uni-bonn.de (JP-A)} \and Sternberg Astronomical
Institute, Moscow State University, Universitetskij Pr. 13, Moscow
119992, Russia\\\email{vgvaram@mx.iki.rssi.ru (VVG)} \and Isaac
Newton Institute of Chile, Moscow Branch,
Universitetskij Pr. 13, Moscow 119992, Russia\\
}

\date{Received 27 April 2010/ Accepted 26 May 2010}

\abstract{The origin of massive field stars in the Large
Magellanic Cloud (LMC) has long been an enigma. The recent
measurements of large offsets ($\sim 100 \, \kms$) between the
heliocentric radial velocities of some very massive (O2-type)
field stars and the systemic LMC velocity provides a possible
explanation of this enigma and suggests that the field stars are
runaway stars ejected from their birthplaces at the very beginning
of their parent cluster's dynamical evolution. A straightforward
way to prove this explanation is to measure the proper motions of
the field stars and to show that they are moving away from one of
the nearby star clusters or OB associations. This approach is,
however, complicated by the long distance to the LMC, which makes
accurate proper motion measurements difficult. We used an
alternative approach for solving the problem (first applied for
Galactic field stars), based on the search for bow shocks produced
by runaway stars. The geometry of detected bow shocks would allow
us to infer the direction of stellar motion, thereby determining
their possible parent clusters. In this paper we present the
results of a search for bow shocks around six massive field stars
that have been proposed as candidate runaway stars. Using archival
{\it Spitzer Space Telescope} data, we found a bow shock
associated with one of our programme stars, the O2\,V((f*)) star
BI\,237, which is the first-ever detection of bow shocks in the
LMC. Orientation of the bow shock suggests that BI\,237 was
ejected from the OB association LH\,82 (located at $\simeq 120$ pc
in projection from the star). A by-product of our search is the
detection of bow shocks generated by four OB stars in the field of
the LMC and an arc-like structure attached to the candidate
luminous blue variable R81 (HD 269128). The geometry of two of
these bow shocks is consistent with the possibility that their
associated stars were ejected from the 30\,Doradus star-forming
complex. We discuss implications of our findings for the problem
of the origin of runaway stars and the early dynamical evolution
of star clusters.}


\maketitle


\section{Introduction}
%

Although the majority of massive stars are situated in their
parent clusters and OB associations, a significant population of
young massive stars exists in the field, some of which are
separated by hundreds of parsecs from known clusters and OB
associations (Garmany, Conti \& Chiosi \cite{ga82}; Garmany
\cite{ga90}; Massey \& Conti \cite{ma83}; Massey et al.
\cite{ma95}). Some Galactic field stars have high measured
peculiar (either radial or transverse) velocities and are
therefore most likely runaway stars ejected from a cluster (Blaauw
\cite{bl61}; Gies \& Bolton \cite{gi86}; Stone \cite{st91};
Zinnecker \cite{zi03}). A straightforward way to prove the runaway
nature of the field OB stars is to use the available kinematic
data on these stars to back-trace their orbits to parent clusters
(e.g. Hoogerwerf, de Bruijne \& de Zeeuw \cite{ho01}). Schilbach
\& R\"{o}ser (\cite{sc08}) make extensive use of this approach to
show that most Galactic field OB stars are formed in clusters.
Alternatively, the runaway nature of the field OB stars can be
proved via detection of their bow shocks -- the natural attributes
of supersonically moving objects (e.g. Baranov, Krasnobaev \&
Kulikovskii \cite{ba71}; Van Buren \& McCray \cite{va88}). The
geometry of detected bow shocks would allow one to infer the
direction of stellar motion (Van Buren, Noriega-Crespo \& Dgani
\cite{va95}), thereby determining the possible parent clusters
even for those field OB stars whose proper motions are still not
available or measured with a low significance (Gvaramadze \&
Bomans \cite{gv08b}; Gvaramadze et al. \cite{gvar10b}). It is
therefore tempting to search for bow shocks around field OB stars
in the Large Magellanic Cloud (LMC) where accurate proper motion
measurements are difficult, while bow shocks can still be resolved
with modern infrared telescopes.

In this paper we present the results of a search for bow shocks in
the LMC using archival {\it Spitzer Space Telescope} data. Our
prime goal was to detect bow shocks produced by isolated, very
massive stars that have previously been qualified as runaways on
the basis of their large peculiar radial velocities (Sect.\,2). We
discovered a bow shock associated with one of these stars, the
O2\,V((f*)) star \object{BI\,237}. A by-product of our search is
detection of bow shocks produced by several other isolated OB
stars, and two of these stars are located around the 30\,Doradus
nebula (Sect.\,3). Implications of our findings for the problem of
the origin of runaway stars and the early dynamical evolution of
star clusters are discussed in Sect.\,4. We use a distance of 50
kpc for the LMC (Gibson \cite{gi00}) so that $1\arcmin$
corresponds to $\simeq 14$ pc.

\section{Very massive field stars as runaways}

The study of the massive star population in the LMC by Massey et
al. (\cite{ma95}) has shown that a large number of young very
massive (O2-type) stars is located at $\sim 100-200$ pc in
projection from star clusters and OB associations. This finding
was interpreted as indicating that the field can produce stars as
massive as those born in clusters (Massey et al. \cite{ma95}). An
obvious alternative to this interpretation is that the massive
field stars were actually formed in a clustered environment and
subsequently ejected from their birth sites via dynamical
processes (e.g. Clarke \& Pringle \cite{cl92}; Pflamm-Altenburg \&
Kroupa \cite{pf06}); i.e., the field massive stars are runaway
stars (Walborn et al. \cite{wa02}; Brandl et al. \cite{br07}). The
large separations from the possible parent clusters and the young
($\sim 2$ Myr) ages of the very massive field stars imply that
their (transverse) velocities should be as high as $\sim 50-100 \,
\kms$ (Walborn et al. \cite{wa02}). The large offsets from the
parent clusters and the high peculiar velocities are not unusual
for Galactic massive runaway stars. For example, the O4\,If star
\object{BD+43$\degr$ 3654} ejected from the Cyg\,OB2 association
is located at about 80 pc from the core of the association
(Comer\'{o}n \& Pasquali \cite{co07}; Gvaramadze \& Bomans
\cite{gv08a}), while the heliocentric radial velocity of the star
is offset by $56 \, \kms$ from the systemic velocity of Cyg\,OB2
(Kobulnicky, Gilbert \& Kiminki \cite{ko10}), which with the
transverse peculiar velocity of the star of $\simeq 40 \, \kms$
(Comer\'{o}n \& Pasquali \cite{co07}) corresponds to the total
peculiar velocity of $\simeq 70 \, \kms$.

The runaway interpretation of the massive field stars in the LMC
received strong support after the discovery that some of them have
high peculiar radial velocities, much greater than the systemic
velocity of the LMC (Massey et al. \cite{ma05}; Evans et al.
\cite{ev06}, \cite{ev10}). Earlier, Nota et al. (\cite{no94}) and
Danforth \& Chu (\cite{da01}) had found that the systemic velocity
of the candidate luminous blue variable S119 (HD\,269687) and its
circumstellar nebula is $\sim 100 \, \kms$ lower than that of the
LMC, so they suggested that S119 could be a runaway star.

To prove the runaway nature of the field massive stars in the LMC
unambiguously, it is necessary to ascertain their parent clusters.
The proper motion measurements cannot help solve this problem in
the near future. At the distance of the LMC, the transverse
velocity of $\sim 50-100 \, \kms$ corresponds to a proper motion
of $\sim 0.2-0.4 \, {\rm mas} \, {\rm yr}^{-1}$, which is too low
to be measured with high confidence using the ground-based
observations. The simpler solution is to infer the direction of
stellar motion via the geometry of bow shocks produced by runaway
stars (Gvaramadze \& Bomans \cite{gv08b}) or, in the case of
evolved massive runaway stars, through the asymmetry in the
brightness distribution of associated circumstellar nebulae
(Danforth \& Chu \cite{da01}; Gvaramadze et al. \cite{gvar09}). It
is worth noting that two very massive Galactic runaway stars,
BD+43$\degr$\,3654 and \object{$\lambda$\,Cep} (O6\,I(n)f; Walborn
\cite{wa73}), are both associated with spectacular bow shocks
(Comer\'{o}n \& Pasquali \cite{co07}; Van Buren et al.
\cite{va95}). One can therefore expect that some of the field
stars in the LMC will manifest themselves in such secondary
attributes of high-velocity runaway stars.

\section{Search for bow shocks in the LMC}

To search for bow shocks, we selected four isolated massive stars
with high peculiar radial velocities (Massey et al. \cite{ma05};
Evans et al. \cite{ev06}, \cite{ev10}) and added two isolated
O2-type stars, \object{BI\,253} (O2\,V((f*))) and
\object{Sk\,$-$68$\degr$137} (O2\,III(f*)), which Walborn et al.
(\cite{wa02}) suggests are runaways owing to their large
separation from their plausible birthplace in the central cluster,
R136 (\object{HD\,38268}), of the 30\,Doradus nebula. The details
of these stars (listed in order of their RA) are summarized in
Table\,\ref{tab:run}. For the first four stars, we give their
peculiar radial velocities, while the transverse velocities are
listed for the remaining two, inferred under the assumption that
both stars were ejected $\sim 2$ Myr ago from R136 (Walborn et al.
\cite{wa02}).

\begin{table}
  \caption{Summary of candidate runaway stars in the LMC.}
  \label{tab:phot}
  \renewcommand{\footnoterule}{}
\begin{tabular}{lccc}
\hline \hline
Star & Spectral type & $v$ ($\kms$) \\
\hline
\object{N11-026} & O2.5\,III(f*) & $\simeq 35^{(a)}$ \\
\object{Sk\,$-$67$\degr$22} & O2\,If* & $\simeq 150^{(b)}$ \\
\object{BI\,237} & O2\,V((f*)) & $\simeq 120^{(b)}$ \\
\object{30\,Dor\,016} & O2 III-If* & $\simeq 85^{(c)}$ \\
\object{BI\,253} & O2\,V((f*)) & $\sim 55^{(d)}$ \\
\object{Sk\,$-$68$\degr$137} & O2\,III(f*) & $\sim 100^{(d)}$ \\
\hline
\end{tabular}
\tablefoot{
 \tablefoottext{a}{Evans et al. \cite{ev06}.}
 \tablefoottext{b}{Massey et al. \cite{ma05}.}
 \tablefoottext{c}{Evans et al. \cite{ev10}.}
 \tablefoottext{d}{Walborn et al. \cite{wa02}.}}
 \label{tab:run}
\end{table}

From our experience in the search for bow shocks produced by OB
stars ejected from Galactic star clusters (Gvaramadze \& Bomans
\cite{gv08a},b; Gvaramadze et al. \cite{gvar10b}) using the
archival data from the {\it Midcourse Space Experiment (MSX)}
satellite (Price et al. \cite{pr01}) and the {\it Spitzer Space
Telescope} (Werner et al. \cite{we04}), we know that the bow
shocks are visible mostly in $21.3\,\mu$m ({\it MSX} band E)
images and $24\,\mu$m images obtained with the Multiband Imaging
Photometer for {\it Spitzer} (MIPS; Rieke et al. \cite{ri04}). The
resolution of {\it Spitzer} $24 \, \mu$m images ($\sim 6\arcsec$)
is three times better than those of the {\it MSX}, so that in the
search for bow shocks in the LMC we utilized the MIPS data alone.

The typical (transverse) size of bow shocks generated by Galactic
OB stars (i.e. the extent of a bow shock in the direction
perpendicular to the vector of the stellar motion) is several
parsecs; e.g. the size of the bow shocks associated with the
above-mentioned massive runaway stars, BD+43$\degr$\,3654 and
$\lambda$\,Cep, is $\simeq 5.0$ and 2.3 pc, respectively. If
placed at the distance of the LMC, these bow shocks will have an
angular size of $\simeq 10\arcsec-20\arcsec$, which is comparable
to or several times greater than the angular resolution of the
MIPS $24\,\mu$m images. Thus, the bow shocks in the LMC can be
resolved with the {\it Spitzer} imaging data!

Visual inspection of MIPS $24\,\mu$m images\footnote{The images,
obtained in the framework of the {\it Spitzer} Survey of the Large
Magellanic Cloud (Meixner et al. \cite{me06}), were retrieved from
the NASA/IPAC Infrared Science Archive
(http://irsa.ipac.caltech.edu).} of fields containing our
programme stars revealed a bow shock associated with only one of
them, namely BI\,237. The non-detection of bow shocks around the
remaining five programme stars is consistent with the
observational fact that only a small fraction ($\la 20$ per cent)
of runaway OB stars produce (observable) bow shocks (Gvaramadze \&
Bomans \cite{gv08b} and references therein).

\begin{figure}
\includegraphics[width=9cm]{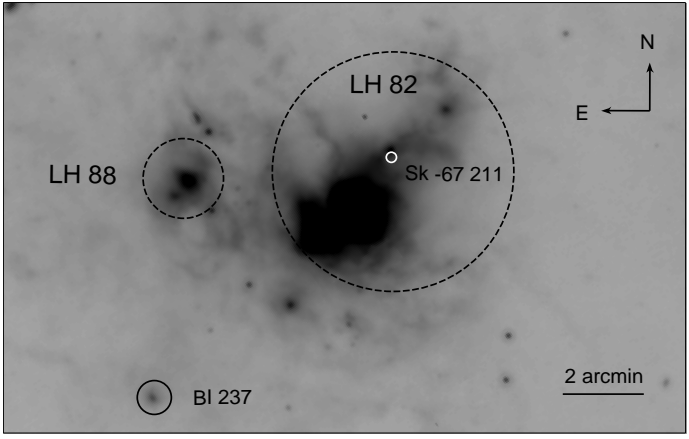}
\centering \caption{MIPS $24\,\mu$m image of the associations
LH\,82 and LH\,88 (indicated by dashed circles). The position of
the O2V((f*)) star BI\,237 and its bow shock are marked by a black
solid circle, while the position of the O2\,III(f*) star
Sk\,$-$67$\degr$211 in LH\,82 is indicated by a white circle. At
the distance of the LMC, $1\arcmin$ corresponds to $\simeq 14$
pc.} \label{fig:BI237}
\end{figure}
\begin{figure}
\includegraphics[width=9cm]{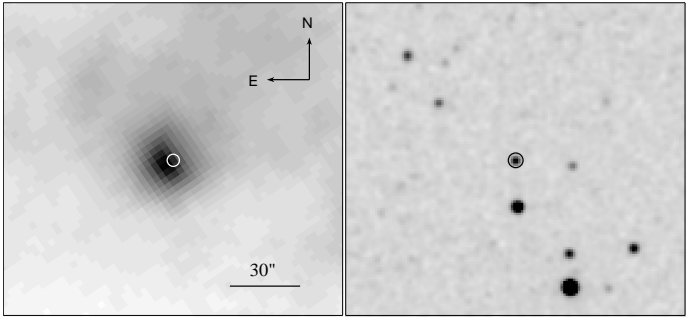}
\centering \caption{{\it Left}: MIPS $24\,\mu$m image of the bow
shock associated with the O2V((f*)) star BI\,237. The position of
BI\,237 is marked by a circle. {\it Right}: 2MASS J band image of
the same field.} \label{fig:BI237bow}
\end{figure}

Figure\,\ref{fig:BI237} gives an overview of the region northwest
of BI\,237 with two associations, \object{LH\,88} and
\object{LH\,82}, whose centres are separated in projection by
$\simeq 65$ and 120 pc from the star. (The approximate boundaries
of the associations are indicated by dashed circles; Bica et al.
\cite{bi99}.) The orientation of the bow shock generated by
BI\,237 (see Fig.\,\ref{fig:BI237bow}) suggests that the more
likely parent association of the star is LH\,82. LH\,82 contains
another very massive star, \object{Sk\,$-$67$\degr$ 211}
(O2\,III(f*); Walborn et al. \cite{wa04}). Assuming that BI\,237
was indeed ejected from LH\,82 and given the young ($\sim 2$ Myr)
age of the star, one finds that its transverse velocity should be
$\sim 60 \, \kms$ if the star escaped from the core of LH\,82 soon
after the birth or higher if the ejection event occurred later on,
so that the total peculiar velocity of the star is $\geq 130 \,
\kms$. The angular size of the bow shock of $\simeq 20\arcsec$
corresponds to the linear size of $\simeq 4.8$ pc, i.e., a figure
typical of massive runaway stars (see above). Using these
estimates, one can constrain the number density of the ambient
interstellar medium, $n_{\rm ISM}$. For the characteristic
(transverse) size of a (parabolic) bow shock of $\sim 2\sqrt{2}
R_0$, where $R_0 =(\dot{M} v_\infty /4\pi \rho _{\rm ISM}
v_{\star} ^2 )^{1/2}$ is the stand-off distance of the bow shock,
$\dot{M}$ and $v_\infty$ are the mass-loss rate and terminal
velocity of the stellar wind, $\rho _{\rm ISM} =1.4m_{\rm H}
n_{\rm ISM}$ is the density of the interstellar medium, $m_{\rm
H}$ the mass of the hydrogen atom, and $v_{\ast}$ ($\geq 130 \,
\kms$) the velocity of the star relative to the ambient medium,
and using the wind parameters of BI\,237, $\dot{M} =7.8\times
10^{-7} \, \msun \, {\rm yr}^{-1}$ and $v_\infty =3400 \, \kms$
(Mokiem et al. \cite{mo07}), one finds $n_{\rm ISM} \leq 0.1 \,
{\rm cm}^{-3}$, i.e., a reasonable number.

A by-product of our search is the detection of bow shocks
associated with four other OB stars in the field of the LMC
(Fig\,\ref{fig:4bow}). The details of these stars are given in
Table\,{\ref{tab:bow}. The spectral types of the stars were found
using the VizieR catalogue access
tool\footnote{http://webviz.u-strasbg.fr/viz-bin/VizieR}. The last
column gives the possible birthplaces of the stars.
\begin{table}
  \caption{Details of bow shock-producing stars.}
  \label{tab:phot}
  \renewcommand{\footnoterule}{}
\begin{tabular}{lccc}
\hline \hline
Star & Spectral type & Association \\
\hline
\object{Sk\,$-$66$\degr$ 16} & O9.7\,Ib$^{(a)}$ & KMHK\,268 \\
\object{Sk\,$-$68$\degr$ 86} & B1:$^{(b)}$ & [SL63] 495  \\
\object{Sk\,$-$69$\degr$ 206} & B2:$^{(b)}$ & 30\,Doradus \\
\object{Sk\,$-$69$\degr$ 288} & B0.5$^{(b)}$ & 30\,Doradus ? \\
\hline
\end{tabular}
\tablefoot{
\tablefoottext{a}{Evans et al. \cite{ev06}.}
\tablefoottext{b}{Rousseau et al. \cite{ro78}}. } \label{tab:bow}
\end{table}
\begin{figure}
\includegraphics[width=9cm]{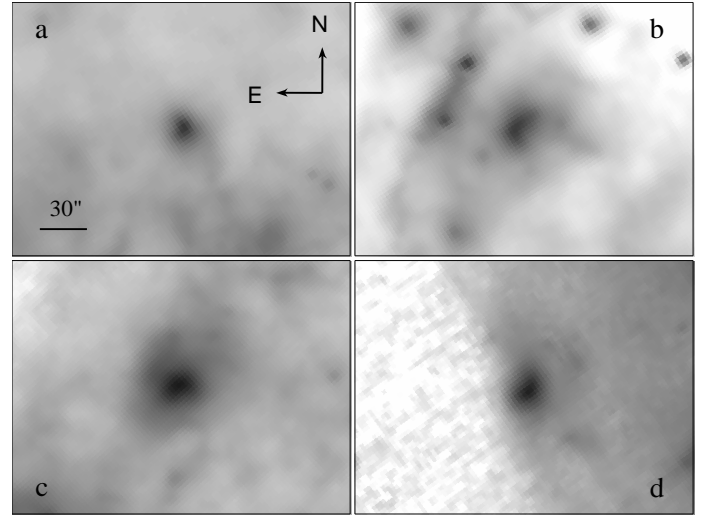}
\centering \caption{MIPS $24\,\mu$m images of bow shocks
associated with (a) Sk\,$-$69$\degr$ 206, (b) Sk\,$-$69$\degr$
288, (c) Sk\,$-$66$\degr$ 16, and (d) Sk\,$-$68$\degr$ 86. The
orientation and the scale of the images are the same.}
\label{fig:4bow}
\end{figure}
\begin{figure*}
\includegraphics[width=18.5cm]{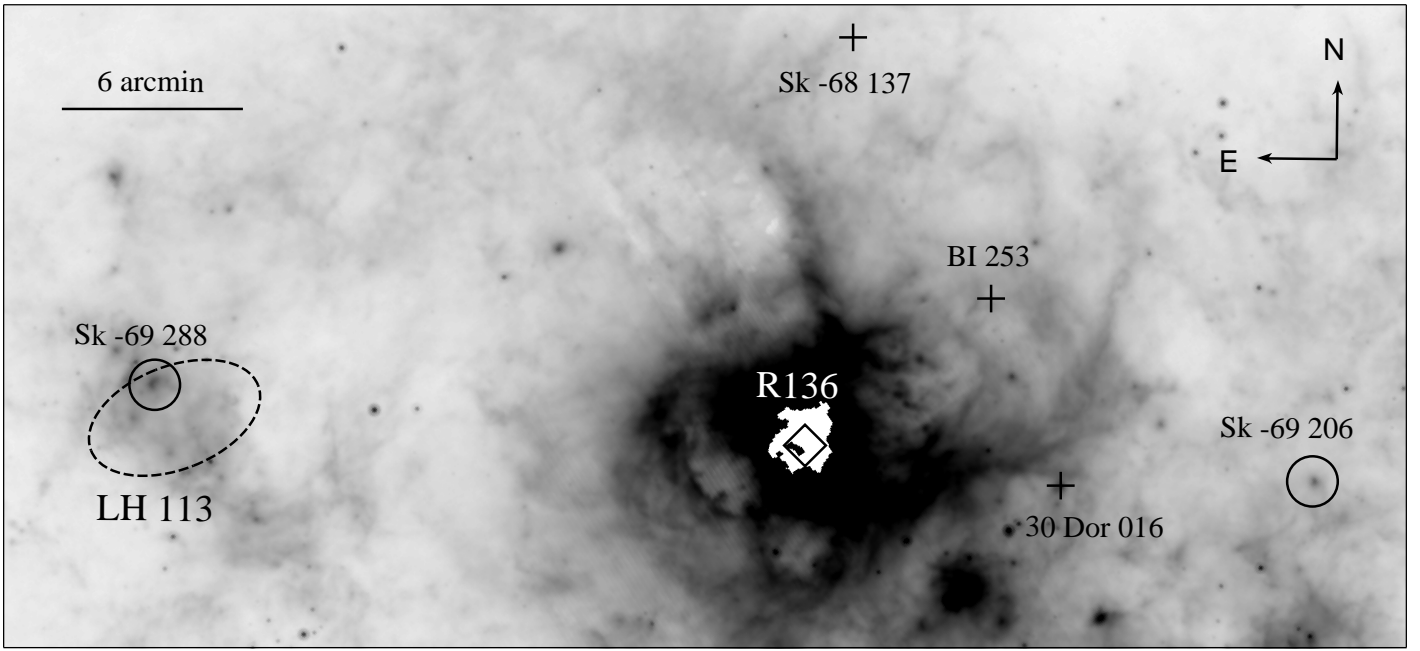}
\centering \caption{MIPS $24\,\mu$m image of the 30\,Doradus
nebula and its surroundings. The positions of Sk\,$-$69$\degr$ 206
and Sk\,$-$69$\degr$ 288 and their bow shocks are marked by solid
circles. The positions of the three programme stars,
Sk\,$-$68$\degr$ 137, BI\,253, and 30\,Dor\,016, are marked by
crosses. The diamond point shows the position of R136. The dashed
ellipsoid indicates the approximate boundary of the association
LH\,113.} \label{fig:30Dor}
\end{figure*}

The bow shock produced by Sk\,$-$69$\degr$ 206 is located at
$\simeq 17\arcmin$ ($\simeq 240$ pc) to the west of R136 -- the
central cluster of the \object{30\,Doradus} nebula
(Fig.\,\ref{fig:30Dor}). The orientation of the bow shock is
consistent with the possibility that the associated star was
ejected from the 30\,Doradus nebula. Assuming that the ejection
event occurred $\sim 2$ Myr ago, one finds the transverse velocity
of Sk\,$-$69$\degr$ 206 of $120 \, \kms$, which is comparable to
the peculiar velocities of the candidate runaway stars in the LMC
(see Table\,\ref{tab:run}).

The bow shock produced by Sk\,$-$69$\degr$ 288 is situated (at
least in projection) within the association LH\,113. The geometry
of the bow shock suggests that Sk\,$-$69$\degr$ 288 is moving away
from the 30\,Doradus nebula, which is located at $\simeq
21\arcmin$ ($\simeq 300$ pc) to the west of the star
(Fig.\,\ref{fig:30Dor}). Although we have no arguments against the
association between Sk\,$-$69$\degr$ 288 and LH\,113, one cannot
exclude the possibility that the actual birthplace of the star is
the 30\,Doradus nebula. In this connection, it is worth noting
that the O5\,V star ALS\,19631 (Hanson \cite{ha03}) was suggested
as a member of the Cyg\,OB2 association on the basis of its
location within the confines of the association (Comer\'{o}n et
al. \cite{co02}). The astrometric data on \object{ALS\,19631} and
the geometry of the bow shock generated by this star, however,
suggest that this runaway was instead ejected from the open
cluster \object{NGC6913} centred $\simeq 3\fdg4$ west of the star
(Gvaramadze \& Bomans \cite{gv08a}).

The bow shocks associated with Sk\,$-$66$\degr$ 16 and
Sk\,$-$68$\degr$ 86 are located, respectively, at $\sim
3.5\arcmin$ ($\simeq 50$ pc) and $\simeq 2.5\arcmin$ ($\simeq 35$
pc) from the clusters \object{KMHK\,268} (Fig.\,\ref{fig:N11}) and
\object{[SL63]\,495} (Fig.\,\ref{fig:Sk}). Sk\,$-$66$\degr$ 16 is
located in the N11 star-forming region, not far from our programme
star N11-026 (see Fig.\,\ref{fig:N11}). We note the detection of a
bow shock-like structure associated with one of the most massive
stars in N11, the ON2\,IIIf*: star \object{N11-031} (Evans et al.
\cite{ev06}). This structure is facing towards the centre of the
parent association \object{LH\,10} (Fig.\,\ref{fig:N11+R81}a).
Interestingly, the radial velocity of N11-031 is $\simeq 30 \,
\kms$ greater than the median velocity of stars in N11 (Evans et
al. \cite{ev06}), which could be considered as indicating that
this star is a runaway as well.

\begin{figure}
\includegraphics[width=9cm]{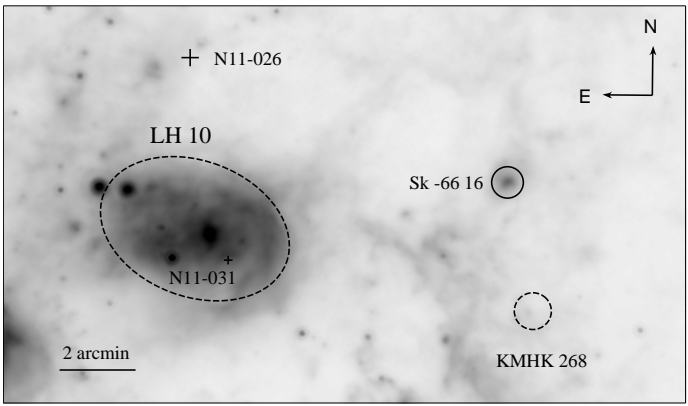}
\centering \caption{MIPS $24\,\mu$m image of the N11 star-forming
region with the programme star N11-026 marked by a large cross.
The small cross indicates the position of the ON2\,IIIf*: star
N11-031. The solid circle shows the position of Sk\,$-$66$\degr$
16 and its bow shock. The positions of the association LH\,10 and
the cluster KMHK\,268 are indicated by a dashed ellipse and a
dashed circle, respectively.} \label{fig:N11}
\end{figure}
\begin{figure}
\includegraphics[width=9cm]{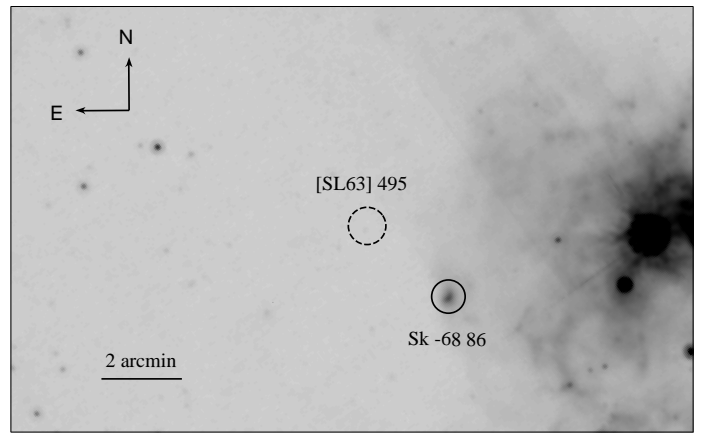}
\centering \caption{MIPS $24\,\mu$m image of the bow shock
associated with Sk\,$-$68$\degr$ 86 (marked by a solid circle).
The position of the cluster [SL63] 495 is indicated by a dashed
circle.} \label{fig:Sk}
\end{figure}

\begin{figure}
\includegraphics[width=9cm]{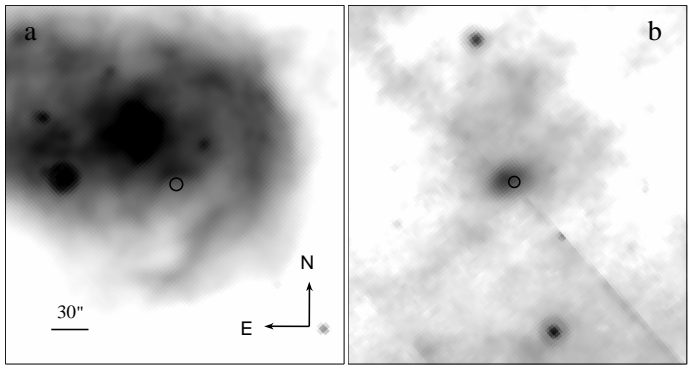}
\centering \caption{MIPS $24\,\mu$m images of (a) a bow shock-like
structure associated with N11-031 and (b) an arc-like nebula
attached to the candidate luminous blue variable R81 (HD\,269128).
The positions of both stars are indicated by circles. The
orientation and the scale of the images are the same.}
\label{fig:N11+R81}
\end{figure}

For the sake of completeness, we note also the detection of an
arc-like nebula (Fig.\,\ref{fig:N11+R81}b) attached to the
candidate luminous blue variable R81 (\object{HD 269128}; Wolf et
al. \cite{wo81}; van Genderen \cite{va01}; cf. Gvaramadze, Kniazev
\& Fabrika \cite{gvar10a}) and the $24 \, \mu$m counterpart to the
circumstellar nebula around the O9f star \object{Sk\,$-$69$\degr$
279} (Weis et al. \cite{we97}).

\section{Discussion and conclusion}

The discovery of a bow shock produced by BI\,237 lends strong
support to the idea that this and other isolated massive stars in
the field of the LMC are runaway stars (Walborn et al.
\cite{wa02}; Massey et al. \cite{ma05}; Evans et al. \cite{ev06},
\cite{ev10}). The young ages ($\sim 2$ Myr) of BI\,237 and other
O2-type field stars decidedly argue for their peculiar velocities
being explained by supernova explosions in binary systems (Blaauw
\cite{bl61}); the massive companion (primary) stars would simply
have no time to end their lives in supernovae. Moreover, the high
(measured or inferred) peculiar velocities of these stars cannot
be accounted for within the framework of the binary-supernova
scenario since it requires that the stellar supernova remnant (a
$5-10 \, \msun$ black hole) receive an unrealistically high ($\ga
200-300 \, \kms$) kick velocity at birth (Gvaramadze \& Bomans
\cite{gv08a}; cf. Gvaramadze \cite{gv09}). The only viable
alternative is that the massive stars were ejected in the field
via dynamical three- or four-body encounters (Poveda et al.
\cite{po67}; Leonard \& Duncan \cite{le90}; Kroupa \cite{kr98};
Pflamm-Altenburg \& Kroupa \cite{pf06}; Gvaramadze, Gualandris \&
Portegues Zwart \cite{gva08}, \cite{gva09}). Naturally, less
massive (late B-type) stars are also ejected from their birth
clusters by dynamical interactions (e.g. Kroupa 1998), but they
would be difficult to observe in the LMC.

The large separation of some of the O2-type field stars from their
plausible birthplaces implies that these stars were ejected soon
after birth (which also argues against the binary-supernova
scenario). This implication has an important consequence for
understanding the early dynamical evolution of star clusters since
it suggests that mass segregation in young clusters (the necessary
condition for effective production of runaway OB stars) should be
primordial rather than caused by the Spitzer instability. For
example, R136 was found to already be mass-segregated at its age
of about 2 Myr or younger (Campbell et al. \cite{ca92}; Hunter et
al. \cite{hu95}; Brandl et al. \cite{br96}; de Grijs et al.
\cite{de02}). But the Spitzer instability could be a very fast
($\la 0.5$ Myr) process if the birth cluster is very dense (e.g.
Kroupa \cite{kr08}). High-precision proper motion measurements for
the massive field stars are therefore required to determine the
timing of their ejections, thereby distinguishing between the
primordial and the dynamical origins of mass segregation in young
clusters (Gvaramadze \& Bomans \cite{gv08b}). Future proper motion
measurements with the space astrometry mission {\it Gaia} will
allow us to solve this problem. At the same time, N-body
experiments are required to quantify the expected differences
between the two types of mass segregation in terms of the ejection
of massive stars.

To conclude, the search for bow shocks in star-forming regions and
subsequent identification of their associated stars serve as a
useful tool for detecting runaway OB stars (e.g. Gvaramadze \&
Bomans \cite{gv08b}; Gvaramadze et al. \cite{gvar10b}), hence for
constraining the dynamical evolution of their parent clusters.
Further search for bow shocks around young massive clusters and OB
associations in the LMC (when necessary, accompanied by follow-up
spectroscopy of their associated stars) is therefore warranted.

\begin{acknowledgements}
We are grateful to S.R\"{o}ser, H.Zinnecker, and the anonymous
referee for carefully reading the manuscript and for useful
comments, allowing us to improve the presentation of the paper.
VVG acknowledges financial support from the Deutsche
Forschungsgemeinschaft. This research has made use of the
NASA/IPAC Infrared Science Archive, which is operated by the Jet
Propulsion Laboratory, California Institute of Technology, under
contract with the National Aeronautics and Space Administration,
the SIMBAD database, and the VizieR catalogue access tool, both
operated at CDS, Strasbourg, France.
\end{acknowledgements}

\end{document}